\documentclass[doublecol]{epl2} 
\usepackage{epstopdf}
\usepackage{subfigure}
\usepackage{graphicx}
\usepackage{amsmath}
\usepackage{xspace}

\title{Measuring deviation from Skumanich braking index in active stars observed by \textit{Kepler} mission}
\shorttitle{Measuring deviation from Skumanich braking index} 

\author{D. B. de Freitas\inst{1}, F. J. Cavalcante\inst{2} and T. M. Santiago\inst{1,3}}
\shortauthor{de Freitas, Cavalcante \& Santiago}

\institute{
\inst{1} Departamento de F\'{\i}sica, Universidade Federal do Cear\'a, Caixa Postal 6030, Campus do Pici, 60455-900 Fortaleza, Cear\'a, Brazil \\
\inst{2}Departamento de F\'{\i}sica, Instituto Federal de Educação, Ci{\^e}ncia e Tecnologia do Cear\'a, Campus Tiangu\'a, 62324-075, Cear\'a,Tiangu\'a, Brazil \\
\inst{3} Centro de Ciências e Tecnologia, Universidade Federal do Cariri, Av. Tenente Raimundo Rocha, Nº 1639, 63048-080, Juazeiro do Norte, Ceará, Brazil.}

\abstract{The aim of this work is to determine the deviation of the value of magnetic braking index $q$ from Skumanich $q=3$ canonical value for giant and main-sequence stars. In this context, the present work attempts to analytically calculate the braking index based on the balance of gravitational and centrifugal forces, a determining factor for understanding the delicate mechanisms that control the spin-down of stars in these evolutionary phases. In the present study, we used a wide sample of stellar targets from the \textit{Kepler} mission with well-defined mass, radius, and rotation period. As a result, \textit{Kepler} stellar parameters provide rather precise values of $q$ index limited in the range $1\leq q\leq 3$, which is consistent with the predictions of the model of magnetic stellar wind. Our results show conclusively that, within the model used in this work, any significant deviation of the braking index away from the value $q=3$ occurs at masses higher than the Kraft break.}

\begin{document}

\maketitle

\section{Introduction}
In the literature, it is widely accepted that the basic mechanism responsible for influencing the loss of angular momentum in low-mass stars (limited in the range between 0.4$M_{\odot}$ and 4$M_{\odot}$) is the magnetic braking. This mechanism combines magnetic wind and a stellar magnetic field. The braking process occurs for stars of different luminosity classes (p. ex., giant and main-sequence stars) but has more interesting results for solar-main-sequence type stars. This process was initially proposed by Schatzman \cite{Schatzman} in 1962. He was the first to point out that the transition between stars with deep envelopes in radiative equilibrium and those with zones of well-developed surface convection occurred between early spectral F-types. In his paper, he explains when mass is ejected from the surface of a star that has a magnetic field, the loss of angular momentum of the star per unit mass is quite expressive. Therefore, even a low mass loss can yield a large effect in reducing the rotation velocity of the star.

In the main-sequence stage, there are two rotational regimes distinguished by effective temperature. With temperatures less than 6200 K, cool main-sequence stars are typically slow rotators and their rotation periods are of a few weeks. According to van Saders \& Pinsonneault \cite{sp}, these periods are far slower than those one would expect were the star to conserve its angular momentum throughout its pre-main sequence collapse and contraction, and far slower than observed rotation periods in very young clusters. On the other hand, hot main-sequence stars, with temperatures greater than 6200K, are rapidly rotating and their periods are less than 10 days. Also known as the Kraft break \cite{Kraft}, this transition between slow and rapid rotators occurs at $\sim$1.3$M_{\odot}$. In this regime, the surface convective envelopes become vanishingly thin, and the stars are unable to yield the magnetic winds and drive angular momentum loss \cite{defreitas2014,Silvaetal13}. 

Roughly, the rotational evolution of main-sequence stars is only driven by the angular momentum loss rate through stellar winds \cite{kawaler1988}. For these stars, the structural changes can be neglected. In contrast, these changes become an important contributor when the stars evolve and become red giants. When stars leave the main sequence, their weak cores due to depleted hydrogen contract and the envelope expands, which leads to an increase in the moment of inertia \cite{weber}. Consequently, as the star evolves toward the giant branches, this increase in the moment of inertia would lead to a decrease in the stellar rotation \cite{kris1997}. This scenario is possible because objects above the Kraft break develop convective envelopes on the subgiant branch. 

After Kraft's paper, Skumanich \cite{sku} showed the rotation, magnetic activity, and lithium abundances for Sun-like stars decrease with the square root of age. Since the Ca$^{+}$ emission lines are linearly proportional to the power of the magnetic field at the stellar surface, then it can be inferred that these surface fields are proportional to the angular velocity which decreases according to Skumanich law. This relation, $v\propto t^{-1/2}$, considers that for stars that belong to the main sequence, the loss of angular momentum, $J$, obeys the relation $\dot{J}\propto-\Omega ^{3}$, where $\Omega$ represents the angular velocity. 

Many efforts were made in order to understand the mechanism responsible for the loss of angular momentum. As an example, we can highlight the works by Mestel \cite{mestel1968,mestel1984} and Mestel \& Spruit \cite{mestel1987}. These studies were relevant because they served as inspiration for Kawaler \cite{kawaler1988} to elaborate his theoretical model. He developed a theoretical model which attributed the loss of angular momentum to the fact that the stars eject material. This stellar wind is trapped by the magnetic field lines and then forced to rotate along with the star causing a change in angular momentum and, consequently, the magnetic braking that decreases the stellar rotation.

According to Kawaler \cite{kawaler1988}, the rate of angular momentum loss due to the presence of wind is proportional to $\Omega^{1+4N/3}$. In this relation, $\Omega$ represents the angular velocity, $a$ denotes the dynamo relationship that varies between 1 and 2 for the unsaturated domain and 0 for the saturated domain, and $N$ is a measurement of magnetic field geometry defined between the values 0 to 2, where 0 represents dipole field geometry and 2 a radial field \cite{defreitas2013}. In this same year, Rutten and Pylyser \cite{rp} investigated the change in the moment of inertia of giant stars with masses of about 1.5$M_{\odot}$. The authors concluded that the loss of angular momentum can be presumably due to magnetic braking and suggest that the magnetic braking by Skumanich applied to G-type stars is a fair estimation in the case of giants. Later, Chaboyer et al. \cite{chaboyer1995} understood that this dependence on $\Omega$ should be reevaluated. Thus, the Kawaler parameterization should be modified in which a saturation level of the magnetic field appears in the angular momentum loss law, as Stauffer and Hartmann \cite{sh} and Bouvier et al. \cite{bouvier} also mention for the main pre-sequence stage.

The so-called braking index ($q$), which is a quantity closely related to the stellar spindown, can provide information as regards the solar-type stars' angular momentum loss mechanisms \cite{sku,kawaler1988,defreitas2013}. de Freitas \& de Medeiros \cite{defreitas2013} investigated the distribution of stellar rotation velocity and the famous relationship between the rotation and stellar age developed by Skumanich \cite{sku}. The authors showed that the rate of stellar angular momentum loss defined by $\dot{J}$ depends on the angular velocity $\Omega $ and varies according to the equation $\dot{J}\propto -\Omega^{q}$, where the index $q$ belongs to non-extensive statistical mechanics \cite{tsallis1988}. This approach was, in a pioneering way, proposed by de Freitas \& De Medeiros \cite{defreitas2013} when working with unsaturated F and G field stars limited in age (from 1 to 10 Gyr) and mass (until 2$M_{\odot}$) within the solar neighborhood. For that proposal, the authors used a catalog of $\sim$16000 stars in the main-sequence stage elaborated by Holmberg et al. \cite{holmberg2007}. In that work, de Freitas \& De Medeiros \cite{defreitas2013} proposed a non-extensive approach to the evolution of stellar rotation based on Kawaler's torque \cite{kawaler1988}. 

After that work, de Freitas et al. \cite{defreitasetal15} generalized the Reiners \& Mohantly's torque \cite{reiners2012} using the same non-extensive framework. In both cases, the authors used the $q$-index from the Tsallis formalism as a parameter that describes the level of magnetic braking as a power law of the type, $t^{1/(1-q)}$. In their work, they show that this parameter $q$ is strongly correlated with stellar mass. For the case of the Reiners \& Mohanty \cite{reiners2012} torque model, the authors showed that the unsaturated magnetic field regime the entropy index can be represented by $1+8a/(4N-5)$. In both models, the saturation regime can be re-established in the non-extensive context assuming the limit $q\rightarrow 1$. 

It is widely accepted in the literature that the rotational braking of slow rotators follows Skumanich's law. In contrast, a deviation from this law is expected when stars evolve by changing their moment of inertia and rotational velocity. On this wise, our main aim is to determine the deviation of the value of magnetic braking index $q$ from Skumanich $q=3$ canonical value for giant and main-sequence stars and, thereby, point out the possible physical mechanisms that control the magnetic braking. 


Our paper is organized as follows. In Section 2, we give the main implications of the effects of changes in the moments of inertia for the magnetic braking index and briefly describe different physical mechanisms that could affect stellar rotation. In Section 3, the working sample is shown. In the next Section, we test our set of stellar targets within the framework of our model and investigate the deviation of the braking index from Skumanich value as a function of the stellar mass and temperature. Finally, in the last Section, conclusions are discussed. 

\section{Modeling the stellar magnetic braking index}

First of all,  the present paper is the second one (Paper I: \cite{defreitas2021}) of a series of three papers which brings a new approach to the nonextensive magnetic braking as a determining factor for understanding the delicate mechanisms that control the spin-down of stars in different luminosity classes. By using the equation proposed by \cite{defreitas2021} to describe the behavior of the rotational velocity, we have that: 
\begin{eqnarray}
\label{4}
\dot{\Omega}=-\frac{\Omega_{0}}{\tau}\left(\frac{\Omega}{\Omega_{0}}\right)^{q}, \quad q\geq 1
\end{eqnarray}
where $\tau$ is defined as a characteristic time, $\Omega_{0}$ is the angular velocity at time $t=0$ and $\Omega$ is the velocity at time $t=t_{\rm age}$, i.e., now, and $q$ as mentioned by de Freitas \& De Medeiros \cite{defreitas2013}.

In the literature, the index $q$ is usually assumed to be a constant that measures the efficiency of magnetic braking during the lives of stars. This braking index can be defined using the second derivate of $\Omega$ as a function of time. Considering that $q$ does not depend on the time, it can be written as
\begin{equation}
q=\frac{\ddot{\Omega}\Omega}{\dot{\Omega}^{2}}.  \label{4.1}
\end{equation}

In the scenario which assumes spin-down due to stellar magnetic winds, the $q$-index is mostly 3. For main-sequence stars, eq. (\ref{4}) is more appropriate when the moment of inertia is a constant.  However, the moment of inertia of a star can be affected by either rotation or stellar radius changes during its evolution. (In the present study, the rate of mass loss is neglected). If the moment of inertia is affected by the rotation, we can approximate the star by an incompressible fluid and, therefore, its volume is constant. On the other hand, if the radius changes over time are taken into account, the volume of the star is not constant. 

In this context, we can write the angular momentum loss rate as
\begin{equation}
\label{cap51}
\dot{J}=I\dot{\Omega} +\dot{I}\Omega,
\end{equation}
on the pre-main-sequence and giant phase, the term involving $\dot{I}$ is very important; on the main-sequence, this term is negligible.

\begin{figure}
	\begin{center}
		\includegraphics[width=0.42\textwidth,trim={2.5cm 1.5cm 2.5cm 1.5cm}]{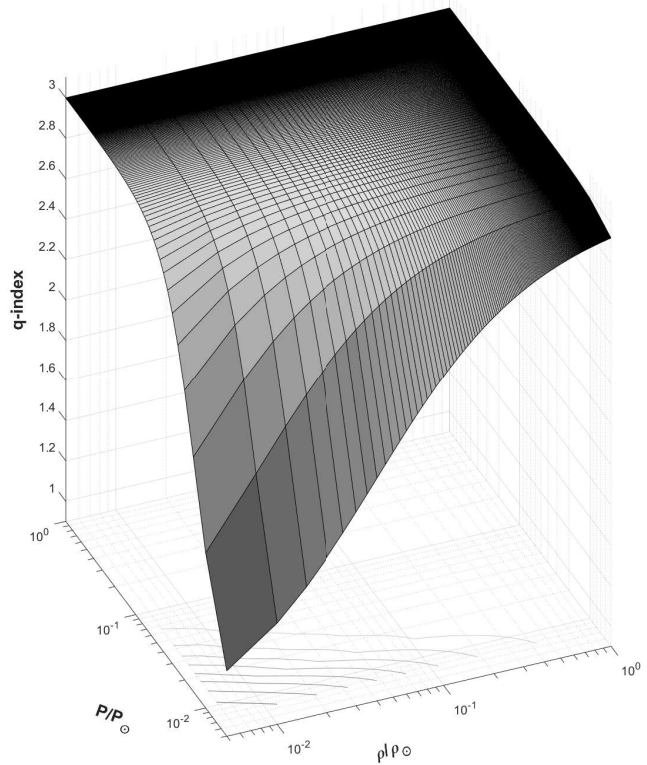}
	\end{center}
	\caption{Behavior of magnetic braking index $q$ as a function of mean density ($\rho/\rho_{\odot}$) and rotation period ($P/P_{\odot}$) as described by eq. (\ref{cap519b1}).}
	\label{fig0}
\end{figure}

\begin{figure}
	\begin{center}
		\includegraphics[width=0.49\textwidth,trim={2.5cm 1.5cm 2.5cm 1.5cm}]{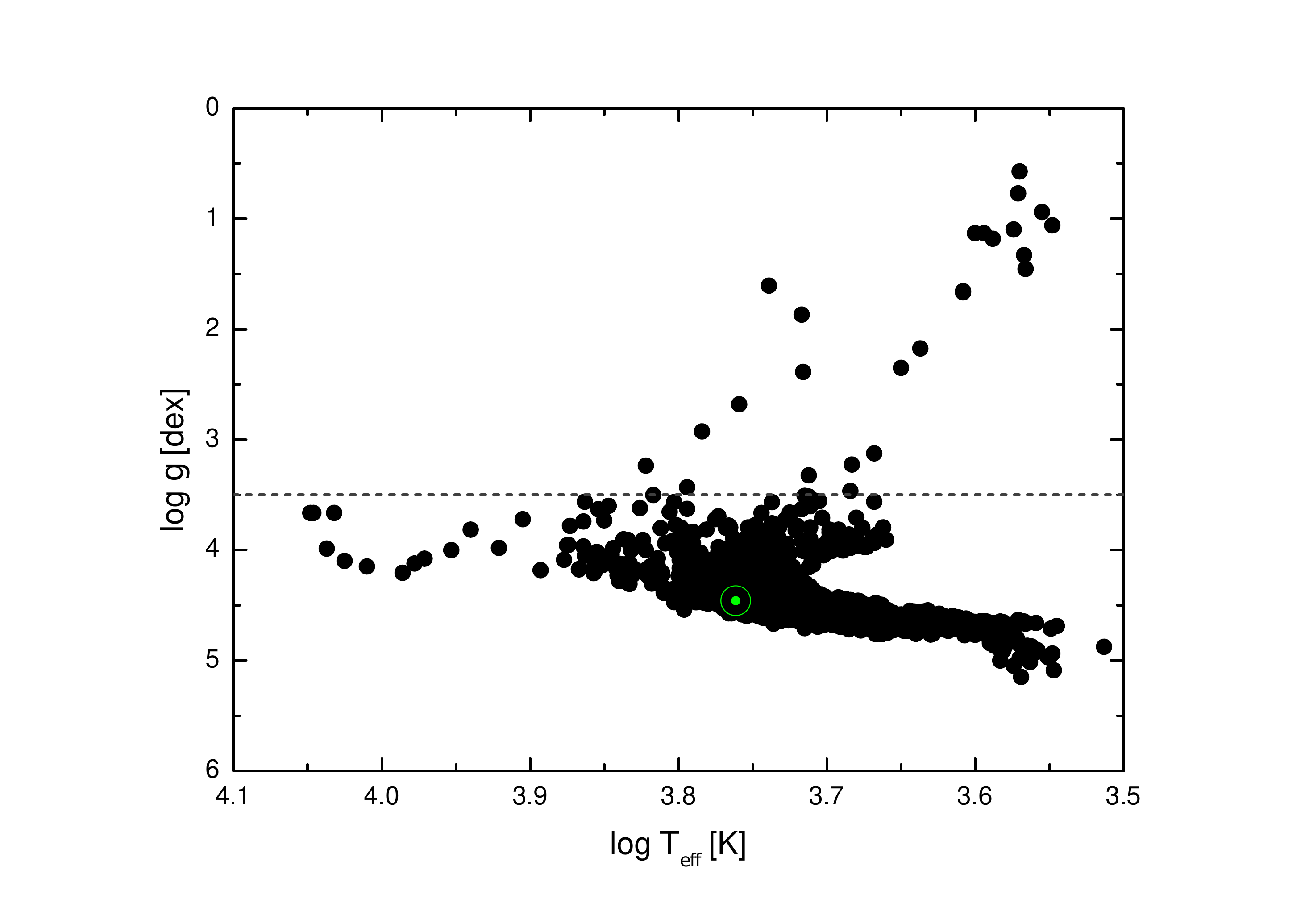}
	\end{center}
	\caption{Effective temperature against effective gravity of all Kepler Q3 stars (black) using KIC parameters. The green star marks the Sun which
is shown for comparison. The dashed line marks $\log g$=3.5 which was set to separate giants in the following.
}
	\label{fig1}
\end{figure}

A star is considered perfectly spherical when its rotational velocity is zero. In fact, rotating stars are considered Maclaurin rotation ellipsoids, where the axis of rotation coincides with the shorter axis passing through its center of mass. A star may remain at a constant volume over time and even exhibit a marked oblateness. In this case, the star is regarded as an incompressible fluid. In general terms, the oblateness depends on the density and the balance of gravitational and centrifugal forces.

From McCoullough's formula, we know that the oblateness can be written as
\begin{equation}
\label{cap52}
\varepsilon=\frac{I-I_{0}}{I_{0}},
\end{equation}
where, $I_{0}$ is the non-rotating spherical moment of inertia and $I$ is the moment of inertia for the rotating star about its axis of rotation. As mentioned by Gizon et al. \cite{gizon}, one mechanism that must be present is rotational oblateness, which is relatively easy to compute when rotation is slow. The centrifugal force distorts the equilibrium structure of a rotating star. The corresponding perturbation of the mode frequencies scales as the ratio of the centrifugal to gravitational forces. Thus, for slow rotators, oblateness is described by a quadrupole distortion of the stellar structure \cite{gizon}. The deviations from spherical symmetry or rotational oblateness $\varepsilon$ depend on the ratio between the centrifugal and gravitational accelerations, as described below
\begin{equation}
\label{cap53}
\varepsilon=\frac{R^{3}\Omega^{2}}{GM},
\end{equation}
where, $\varepsilon$ is proportional to the square of the rotation velocity. Additionally, apart from the gravitational and pressure forces, the gas in the stellar inner is under the action of a centrifugal force. If this force is small, the spherical symmetry is not affected. Therefore, the condition to be fulfilled to maintain the spherical symmetry is $\varepsilon\ll 1$.

Compiling the equations \ref{cap52} e \ref{cap53}, we find that the moment of inertia, as a function of the rotational velocity of the star, is given by
\begin{equation}
\label{cap58}
I=I_{0}\left(1+\frac{R^{3}\Omega^{2}}{GM}\right),
\end{equation}
where $I$ explicitly depends on $\Omega^{2}$. Hence, we have
\begin{equation}
\label{cap510}
\dot{I}=\frac{\mathrm dI}{\mathrm d\Omega}\frac{\mathrm d\Omega}{\mathrm dt},
\end{equation}
where, from eq. \ref{cap58}, we find that $\frac{\mathrm dI}{d\Omega}$ is
\begin{equation}
\label{cap511}
\frac{\mathrm dI}{\mathrm d\Omega}=2I_{0}\frac{R^{3}\Omega}{GM},
\end{equation}
and therefore,
\begin{equation}
\label{cap512}
\dot{I}=2 I_{0}\frac{R^{3}\Omega}{GM} \dot\Omega.
\end{equation}

Thus, the momentum angular loss rate (see eq. \ref{cap51}) can be rewritten as
\begin{equation}
\label{cap514}
\dot{J}=I_{0}\dot{\Omega}\left(1+\frac{3R^{3}\Omega^{2}}{GM}\right).
\end{equation}

Let us assume that the momentum angular loss rate is a function of only the rotational velocity $\Omega$ and decays as a power law given by
\begin{equation}
\label{cap515}
\dot{J}=K_{m}\Omega^{m},
\end{equation}
where $m$ is a parameter which controls the loss rate and $K_{m}$ is a coefficient that is related to a model which describes the angular momentum loss law for the unsaturated magnetic field. We will consider the coefficient $K_{m}$ constant in time. On the other hand, for stars with high rotation rates, the magnetic field must be saturated \cite{reiners2012}, i.e. $m=1$ and, thereby
\begin{equation}
\label{cap515X}
\dot{J}=K\Omega.
\end{equation}

In the literature, there are several parametrizations that describe angular momentum loss through the magnetized wind. Among them, the prescriptions of \cite{kawaler1988}, \cite{reiners2012} and \cite{matt} have been widely tested. In each case, there is a functional form for $K_{m}$. However, in the present study, it is not necessary to define a previous form for $K_{m}$ due to the general nature of our model.	

Combining equations \ref{cap514} and \ref{cap515}, we obtain that the first-order derivative of $\Omega$ is given by
\begin{equation}
\label{cap516}
\dot{\Omega}=\frac{K_{m}}{I_{0}}\left(\frac{\Omega^{m}}{1+\frac{3R^{3}\Omega^{2}}{GM}} \right).
\end{equation}

Similarly, by deriving the above equation, we can find the second-order derivative of $\Omega$ is
\begin{equation}
\label{cap517}
\ddot{\Omega}=\frac{K_{m}}{I_{0}}\left(\frac{\Omega^{m-1}\dot{{\Omega}}}{1+\frac{3R^{3}\Omega^{2}}{GM}}\right)\left(m-\frac{\frac{6R^{3}\Omega^{2}}{GM}}{1+\frac{3R^{3}\Omega^{2}}{GM}} \right).
\end{equation}

So, we already have the first and second derivatives of $\Omega$ (see eq. \ref{4.1}) and, thus, we can find the following expression for the magnetic braking index as a function of $\Omega$, $M$ and $R$:
\begin{equation}
\label{cap518}
q(\Omega,R,M)=m-\frac{\frac{6R^{3}\Omega^{2}}{GM}}{1+\frac{3R^{3}\Omega^{2}}{GM}},
\end{equation}
where, as our interest is to measure the deviation from the Skumanich index, the parameter $m$ must be equal to 3 and, for that reason, eq. (\ref{cap518}) can be rewritten as
\begin{equation}
\label{cap519b}
q(\Omega,R,M)=3-\frac{\frac{6R^{3}\Omega^{2}}{GM}}{1+\frac{3R^{3}\Omega^{2}}{GM}}.
\end{equation}

It is clear from the equation above that non-rotating stars (or slowly-rotating ones) obey the Skumanich relationship \cite{sku}. Otherwise, any combination of $\Omega$, $R$, and $M$ will imply a magnetic braking index $q$ becomes less than 3.  

Based on our sample parameters (see the next Section), we will consider the rotation period $P$ rather than the angular velocity $\Omega$ and mean density $\rho$ instead of $M$ and $R$. Considering $\Omega=2\pi/P$, we can rewrite equation (\ref{cap519b}), in the solar terms, as
\begin{equation}
\label{cap519b1}
q(P,\rho)=3-\frac{2}{1+\frac{1}{3\varepsilon_{\odot}}\left(\frac{\rho}{\rho_{\odot}}\right) \left(\frac{P}{P_{\odot}}\right)^{2}},
\end{equation}
with
\begin{equation}
\label{cap519b2}
\frac{\rho}{\rho_{\odot}}=\left( \frac{M}{M_{\odot}}\right)\left( \frac{R}{R_{\odot}}\right)^{-3},
\end{equation}
where $\rho_{\odot}=1.41$ g/cm$^{3}$ is the solar mean density. The solar oblateness, $\varepsilon_{\odot}$, is $2.14\times 10^{-5}$, assuming $\Omega_{\odot}=2.9\times 10^{-6}$ s$^{-1}$ ($P_{\odot}$=25.4 days), $R_{\odot}=6.96\times 10^{8}$ m, $M_{\odot}=1.99\times 10^{30}$ kg, the gravitational constant $G=6.67\times 10^{-11}$ Nm$^{2}$kg$^{-2}$. Based on these data, the solar $q$-index is the Skumanich value, $q=3$. Equation (\ref{cap519b1}) clearly shows that $q$ is defined between 1 and 3, considering that the term $\left(\frac{\rho}{\rho_{\odot}}\right) \left(\frac{P}{P_{\odot}}\right)^{2}$ strongly depends on the evolutionary stage. Theoretically, for solar-type stars is expected $q=3$ and the effect of $\dot{I}$ can be neglected. Although, mostly for giant stars, when $q\sim1$, $\dot{I}$ is decisive. In contrast, there may be some giant stars where the value of $q$-index is close to the canonical value $q=3$. All these cases will be discussed in Results and Discussions.

\begin{figure}
	\begin{center}
		\includegraphics[width=0.49\textwidth,trim={2.5cm 1.5cm 2.5cm 1.5cm}]{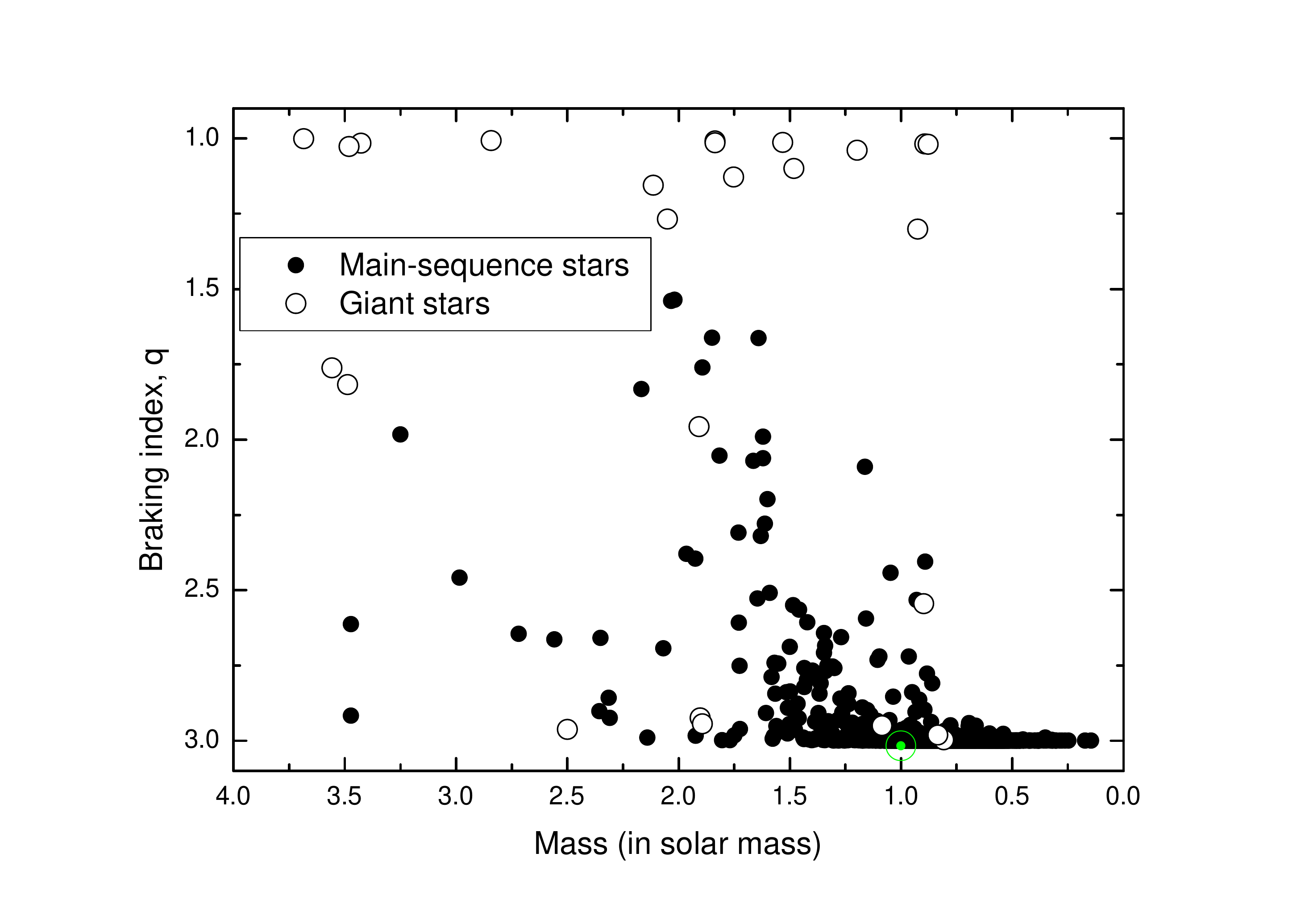}
	\end{center}
	\caption{Braking index $q$ as a function of stellar mass calculated for our final sample of stars.}
	\label{fig2}
\end{figure}

\begin{figure}
	\begin{center}
		\includegraphics[width=0.49\textwidth,trim={2.5cm 1.5cm 2.5cm 1.5cm}]{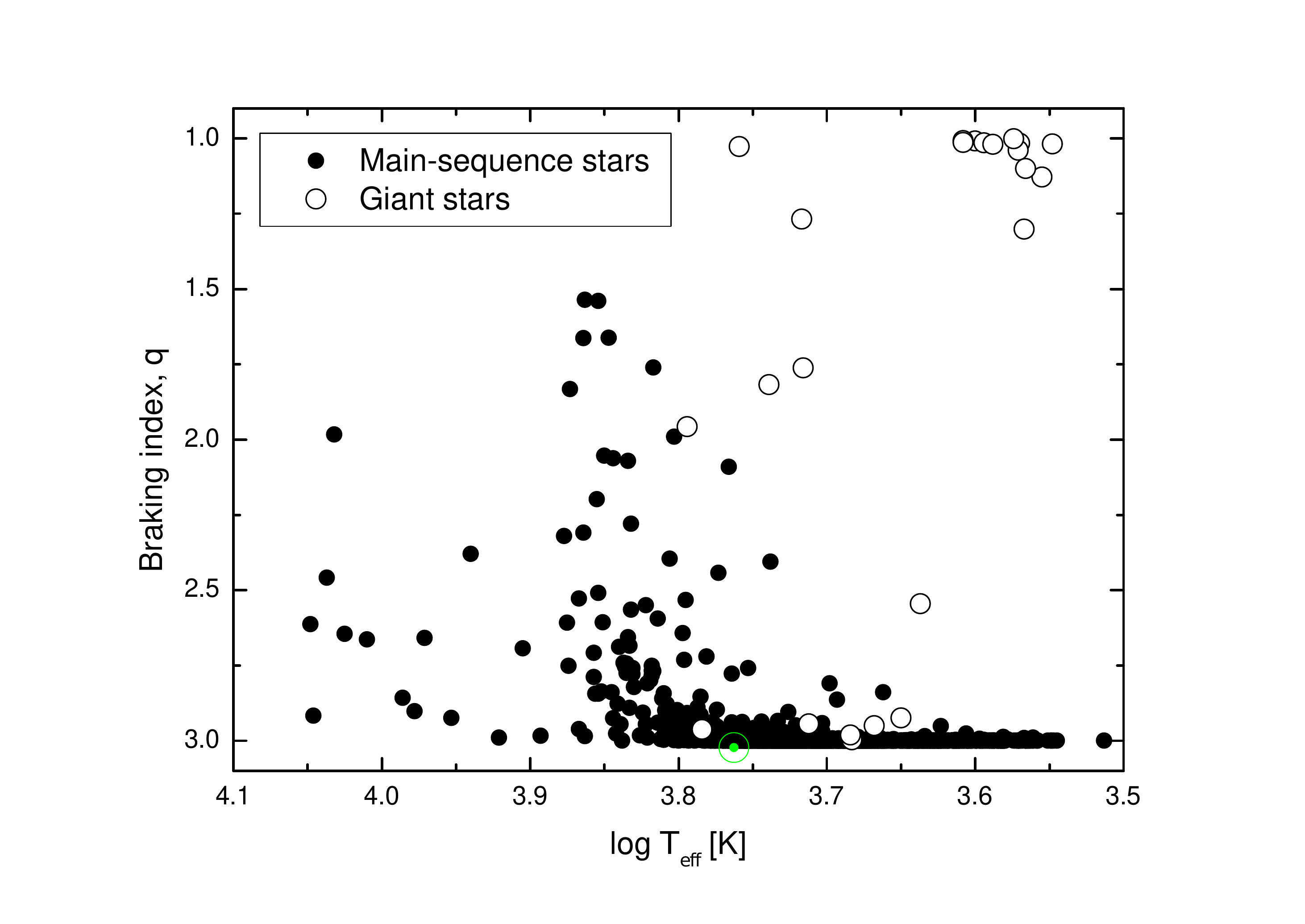}
	\end{center}
	\caption{Braking index $q$ as a function of effective temperature calculated for our final sample of $\sim$ 16,000 stars. As illustrated in Figure \ref{fig1}, the Sun is denoted by green symbol $\odot$.}
	\label{fig3}
\end{figure}

\begin{figure}
	\begin{center}
		\includegraphics[width=0.49\textwidth,trim={2.5cm 1.5cm 2.5cm 1.5cm}]{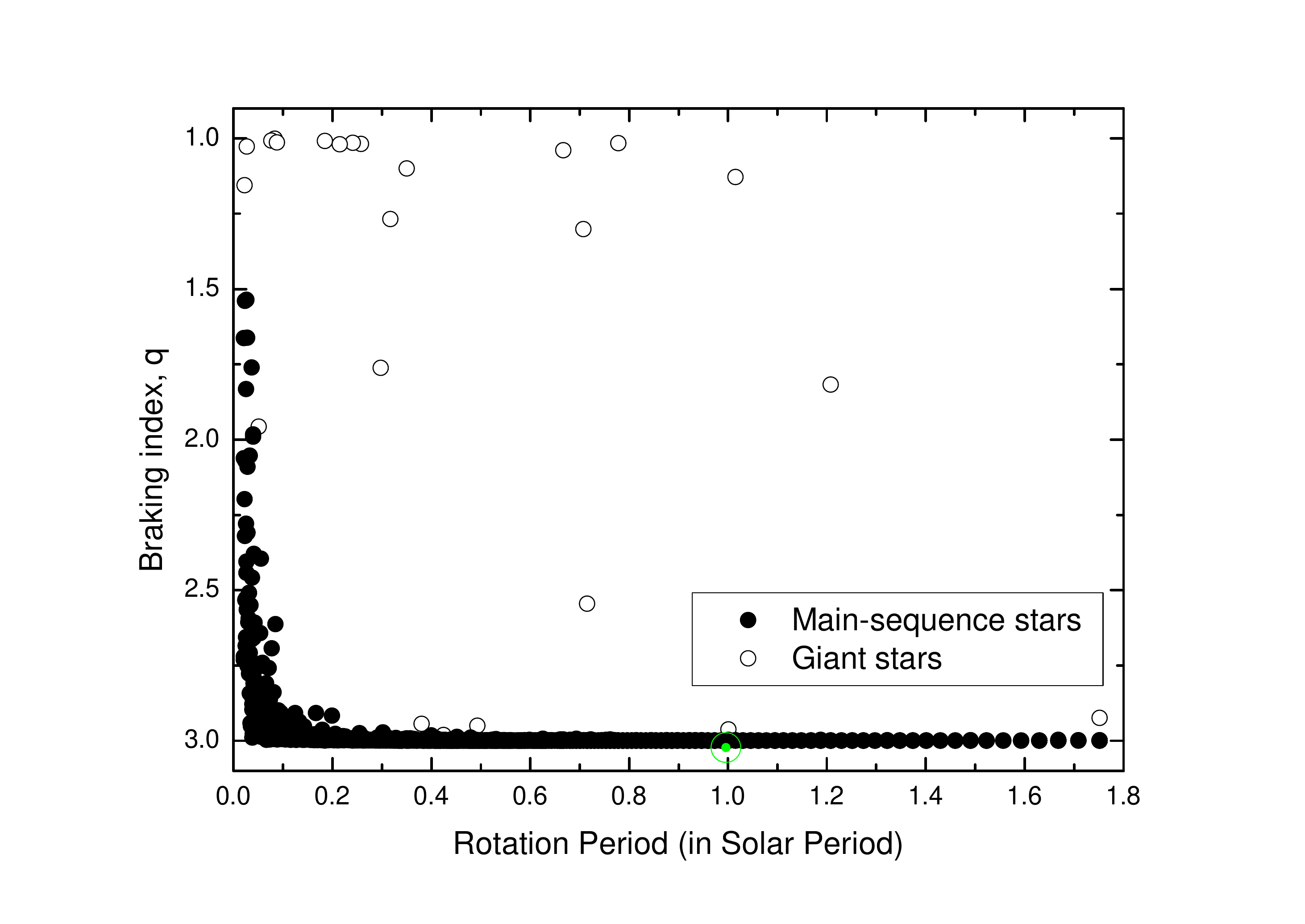}
	\end{center}
	\caption{Braking index $q$ as a function of rotation period calculated for our final sample of stars.}
	\label{fig4}
\end{figure}

According to Figure \ref{fig0}, the index increases more smoothly with increasing density, while it increases faster as the stars rotate more slowly. There is a region outside the contour lines (see the $\rho-P$ plan) where the Skumanich index ($q=3$) is dominant. Only in the small region bounded by these curves does the braking index significantly differ from the canonical Skumanich value. In the region where the stars are less dense than the Sun and hundreds of times faster than it, the index is closer to the value $q=1$. According to de Freitas \& De Medeiros \cite{defreitas2013}, this value is associated with stars that have already saturated the magnetic field.

\section{Working Sample}
Our sample of stars is based on the catalogs by Mathur et al. \cite{mathur2017} and Reinhold et al. \cite{reinhold} both based on the analysis of light curves from the \emph{Kepler} mission. The sample by Mathur et al. \cite{mathur2017} presents revised stellar properties for 197,096 Kepler targets observed between quarters 1-17 (Q1-17). Among the available stellar properties, mean density, mass, radius, and effective temperature are the most important in the present study. However, this sample does not provide us with information about the rotation. 

In this sense, we used the catalog by Reinhold et al. \cite{reinhold} to extract the rotation period from those stars studied by Mathur et al. \cite{mathur2017}. The authors analyzed a sample of 40,661 active stars and determined the rotation period of approximately 24,124 stars with values between 0.5 and 45 days. Unlike the sample by Mathur et al. \cite{mathur2017}, the catalog by Reinhold et al. \cite{reinhold} used only Q3. According to the authors, the Q3 data was chosen because it has fewer instrumental effects than earlier quarters, and carries a large number of targets (165,548 light curves in total). Furthermore, the rotation period is expected to change very little in a time window of just over 4 years.

We combined these two catalogs and obtained a sample of more than 16,000 active stars with temperatures between 3,257 and 11,117 K and mass distributed in the range between 0.2 and 4 solar masses, i.e., low-mass stars. As shown in Figure \ref{fig1}, our sample is divided into two classes of luminosity: giants and main-sequence stars. The number of main-sequence stars is dominant and comprises about 98\% of the sample. Unfortunately, there are only a few giant stars with a well-defined rotation period, but our numbers are enough to understand the behavior of the $q$ index for this class of luminosity.

Getting an idea about the acquisition of Kepler mission data and its instrumental properties, a brief description follows. The \emph{Kepler} mission performed 17 observational runs of $\sim$90 days, each of which was named by Quarters\footnote{\texttt{http://archive.stsci.edu/pub/kepler/lightcurves/tarfiles/}} comprised of long cadence (data sampling every 29.4~min \cite{Jenkins2010}) and short cadence (sampling every 59~s) observations \cite{van,thompson}; detailed discussions of the public archive can be found in many \emph{Kepler} team publications, e.g., \cite{boru2009,boru}, \cite{batalha}, \cite{koch}, and \cite{basri2011}. Regarding data format, the \emph{Kepler} archive provides both Simple Aperture Photometry data (processed using a standard treatment that only removes more relevant spacecraft artifacts) and Pre-Search Data Conditioning (PDC) data that are processed using a refined treatment based on the \emph{Kepler} pipeline \cite{Jenkins2010b}, which removes more thermal and kinematics effects arising from spacecraft operation \cite{van2}.
	
\section{Discussions and Final Remarks}
In this paper, we study the deviation of the value of magnetic braking index $q$ from the Skumanich $q=3$ canonical value for giant and main-sequence stars. We model the braking index of these classes of stars taking into account the effect of a change in the moment of inertia, mean density as well as rotation period. We show that the appropriate combination of these very quantities can account for the braking indices observed for active \textit{Kepler} stars. As a result, their stellar parameters provide rather precise values of $q$ index limited in the range $1\leq q\leq 3$, which is consistent with the predictions of the model of magnetic stellar wind.

Generally speaking, the stellar structure theory predicts that the stellar mean density decreases as the mass increases (from M to O stars) \cite{kipp}. Based on our model, Equation (\ref{cap519b1}) reveals that a greater variation of the magnetic braking index occurs in more massive stars than 1.3$M_{\odot}$ as can be observed in Figure \ref{fig2} for the main-sequence stars. This behavior is expected since more massive stars are fast rotators and, therefore, the effect of a change of the moment of inertia on these more massive stars is more pronounced. Nevertheless, it can be omitted for very low-mass stars (slow rotators) that occupy the bottom right corner of Figure \ref{fig2}. This result can also be confirmed in Figure \ref{fig3} since the stellar parameters mass and effective temperature are equivalent.

As can be seen in Figure \ref{fig4}, the relationship between the index $q$ and the rotation period $P$ allows us to state that fast rotators in the main-sequence ($P/P_{\odot}<0.1$, i.e., stars that rotate at least ten times faster than the Sun) assume values of $q$ within a wide range of 1.5 to 3 at masses higher than the Kraft break as can be verified by Figure \ref{fig2}. While giant stars with high rotation have values of $q$ very close to 1, suggesting that they are close to the saturated magnetic field regime. On the other hand, slower main-sequence stars ($P/P_{\odot}>0.1$) undergo Skumanich's braking law, indicating that only the braking exponent $q=3$ is needed to describe the rotational behavior of the star throughout its main sequence lifetime. However, for the slower giant stars, there is an apparent spread in the values of $q$. Few giant stars undergo Skumanich-type braking. For them, the effect of varying the moment of inertia can be neglected. In the opposite situation, if the index $q$ is less than 3, we can infer that the magnetic braking has been weakened, that is, the component that considers $q(\dot {I}\neq 0)$ intensifies the deviation of the Skumanich braking index \cite{van}. This interpretation also holds for main-sequence stars.

In particular, it is not common to find short-period giant stars. In this case, it is possible they have engulfment a planet, changing its rotation. In fact, this is predicted because of the expansion of the convective envelope. In the case of planet engulfment, there are studies that find an anomalous increase in stellar rotation velocity while the star is still on the main-sequence \cite{ocbs,melendez}. In contrast, the main prediction of our model is that the braking index of low-mass giant and main-sequence stars, without the presence of a binary companion or planet, depends only on density and rotation period. This issue about the influence of planetary engulfment on the braking index will be addressed in a forthcoming communication (Paper III).

Undoubtedly, the strongest conclusion in this study concerns the role of the deviation of Skumanich's magnetic braking as a ``new parameter'' to explain the mechanism that controls stellar rotation. Furthermore, it is sensitive to the oblateness effects that determine how strong magnetized winds are. It also provides us with a powerful diagnostic tool concerning the "rotational lifetime" of the stars on different luminosity classes. In particular, the weakened magnetic braking as the origin of the anomalously rapid rotation, as suggested by Van Clave et al. \cite{van}, is, in the context of our model, understood as an increase in the broadness of the deviation of the Skumanich braking index and, therefore, associated with a change in the moment of inertia of the stars. This is a strong point of our model, in that the it can improve the diagnosis of the transition to weaker magnetized winds.

\acknowledgments
DBdeF acknowledges financial support 
from the Brazilian agency CNPq-PQ2 (Grant No. 305566/2021-0). Research activities of STELLAR TEAM of Federal University of Cear\'a are supported by continuous grants from the Brazilian agency CNPq.

\end{document}